# Low-cost Cognitive Radios against Spectrum Scarcity

Dr. Alexandros-Apostolos A. Boulogeorgos[1], Member, IEEE,
Prof. George K. Karagiannidis[2], Fellow, IEEE

[1]Department of Digital Systems, University of Piraeus, Greece
(e-mail: al.boulogeorgos@ieee.org)

[2]Department of Electrical and Computer Engineering, Aristotle University of Thessaloniki,
Greece (e-mail: geokarag@auth.gr)

## 1. Abstract

The next generation wireless networks are envisioned to deal with the expected thousand-fold increase in total mobile broadband data and the hundred-fold increase in connected devices. In order to provide higher data rates, improved end-to-end performance, low latency, and low energy consumption at low cost per transmission, the fifth generation (5G) systems are required to overcome various handicaps of current cellular networks and wireless links. One of the key handicaps of 5G systems is the performance degradation of the communication link, due to the use of low-cost transceiver in high data rate. Motivated by this in this paper, we discuss the impact of transceiver front-end hardware imperfections on the spectrum sensing performance of cognitive radios.

## 2. Introduction

Radio frequency (RF) wireless spectrum is one of the most tightly regulated communication resources. From the early days of wireless communications, regulatory bodies were concerned about the interference that will be caused by different uses of the wireless spectrum. These concerns lead to the "doctrine of spectrum scarcity", which assigned each piece of spectrum with certain bandwidth to specific wireless systems [1]. With the proliferation of wireless services, in the last couple of decades, in several countries, most of the available spectrum has been fully (or almost-fully) allocated, which results in the spectrum scarcity problem. On the other hand, several studies have revealed that an important amount of spectrum experience low utilization (see e.g., [2] and references therein). Therefore, in order to maintain sustainable development of the wireless communication industry and market, rethinking of the spectrum allocation policies is necessary.

In this context, cognitive radios (CRs) are envisioned as one of the key enablers to deal with the RF spectrum scarcity issue. CRs are intelligent reconfigurable wireless devices capable of sensing the conditions of the surrounding RF environment and modifying their transmission para-meters accordingly for achieving best overall performance without interfering with other users. As a result, CR have recently been adopted in several wireless communication standards, such as long term evolution advanced (LTE-A), wireless fidelity (WiFi-IEEE 802.11), Zigbee (IEEE 802.15.4), and worldwide interoperability for microwave access (WiMAX-IEEE 802.16) [3].

One important task of CRs is spectrum sensing, i.e., the identification of temporarily vacant portions of spectrum. Spectrum sensing allows the exploitation of the under-utilized spectrum; hence, it is considered to be the main countermeasure against the spectrum scarcity problem. Moreover, it is an essential element in the operation of CRs. From technological point of view, in order to enable multiple-frequency band spectrum sensing, radio transceivers need to be flexible and software re-configurable devices. By definition, flexible radios are characterized by the ability to operate over multiple-frequency bands, and to support different types of waveforms, as well as various air interface technologies of currently existing and emerging wireless systems [4]. In this sense, the terms multi-mode, multi-band, and multi-standard are commonly used. The flexibility of transceivers is in-line with the software define radio (SDR) principle, which is considered to be one of the key technologies that enables the use of CRs [5].

From an economical point of view, the advantages in integrated circuit technologies and the adoption of low-complexity transceiver structures, such as the direct-conversion radio (DCR) architecture, allowed improvements in manufacturing efficiency and automation that resulted in reducing the cost-per-device. Moreover, the use of low-complexity transceiver structures enable the reduction of the power consumption in battery-powered devices, without sacrificing too much performance. However, these advantages come with a cost in the device's hardware quality.

### 3. Hardware imperfections in Low-cost CR devices

In general, the demands for multi-standards operation, flexibility in order to deal with the spectrum scarcity problem, and higher data rate, as well as the constraints of product cost, device size, and energy efficiency, lead to the use of simplified radio architectures and low-cost radio electronics [6]. In this context, the DCR architecture provides an attractive front-end solution, since, as illustrated in Fig. 1, it requires neither external intermediate frequency filters nor image rejection filters [7]. Instead, the essential image rejection is achieved through signal processing methods. DCR architectures are low cost and can be easily integrated on-chip, which render them excellent candidates for modern wireless technologies [8]. However, direct-conversion transceivers are typically sensitive to front-end related impairments, such as in-phase (I) and quadrature (Q) imbalance (IQI), local oscillator (LO) phase noise and amplifiers nonlinearities, which are often inevitable due to components imperfections and manufacturing defects. Motivated by this, this section is focused on presenting the impact of hardware imperfections in spectrum sensing.

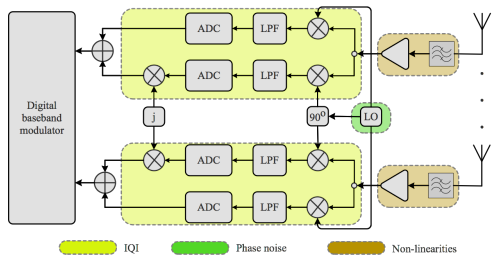

**Fig. 1:** DCR receiver architecture.

*Amplifiers nonlinearties:* Amplifiers nonlinearities cause spurious signals, which play the role of interference in adjacent channels. When the amplifier is used simultaneously by a number of carriers, intermodulation products are generated, which result to distortion in the desired signals. According to Bussgant's theorem, the amplifiers non-linearities results to an amplitude/phase distortion and a nonlinear distortion noise. Fig. 2.b intuitively presents the impact of amplifiers nonlinearities in spectrum sensing. In more detail, we consider a multi-channel spectrum sensing scenario in which the CR decides whether the channel k out of the K channels (in this case K=6) is occupied. We observe that due to the non-linearities, the signal power in an occupied channel is amplified, i.e., if a channel is busy, the accuracy of the correct is increased. On the other hand, if the channel is idle, the nonlinearities cause a noise power increase; hence, the detector can falsely decide that the channel is occupied. As discussed in [9], in order to mitigate the impact of false detection, due to the impact of amplifiers nonlinearities, the CR designer should appropriately adjust the spectrum sensing threshold.

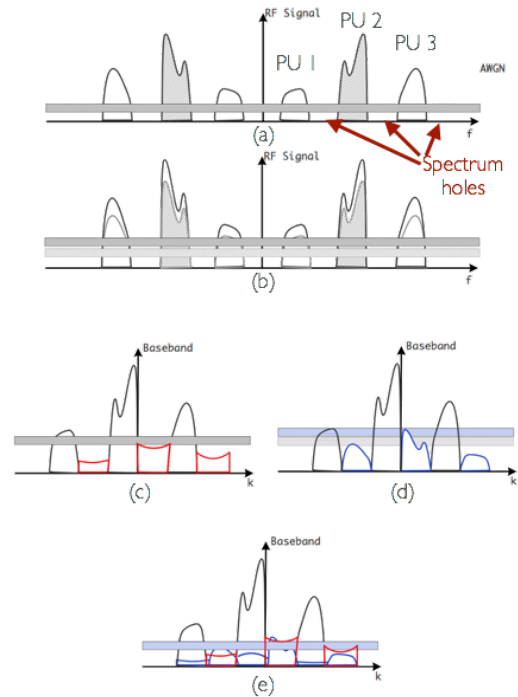

**Fig. 2:** Spectra of the received signal: (a) before the low noise amplifier (LNA) (passband RF signal), (b) after LNA (pass- band RF signal), (c) after down-conversion (baseband signal), when local oscillator's phase noise is considered to be the only RF imperfection, (d) after down-conversion (baseband signal), when IQI is considered to be the only RF imperfection, (e) after down-conversion (baseband signal), the joint effect of LNA nonlinearities, phase noise and IQI. In this figure PU stands for the primary user.

*LO Phase noise*: Noise is of major concern in LOs, because introducing even small noise into a LO leads to dramatic changes in its frequency spectrum and timing properties. This

**IEEE COMSOC TCCN Communications**

phenomenon, peculiar to LOs, is known as phase noise or timing jitter, and it was identified as one of the major performance limiting factors of communication systems in several studies (see for example [9] and references therein). Generally, the disturbance of the amplitude of the oscillator output is marginal. As a result, most influence of the oscillator imperfection is noticeable in random deviation of the frequency of the oscillator output. These frequency deviations are often modelled as a random excess phase, and therefore referred to as phase noise. Phase noise will more and more appear to be a performance limiting factor especially in the case of multi-carrier and multi-channel communications, when low-cost implementations or systems with high carrier frequencies are considered, since, in these cases, it is harder to produce an oscillator with sufficient stability. As illustrated in Fig.2.c, phase noise causes adjacent channel interference in channel k from the channels k-1 and k+1. As a result, an idle channel might be identified as busy, due to the power leakage from a neighbor occupied channel. As illustrated in Fig. 3, this phenomenon can significantly limit the spectrum sensing capabilities of the CR [9], [10]. In more detail, assuming that the signal to noise ratio (SNR) for all the K channel is the same, we observe that, for a fixed SNR, as the 3 dB bandwidth of the LO, $\beta$, increases, the interference from adjacent channel increases, and the spectrum sensing capabilities of the CR decreases in comparison with the corresponding capabilities in ideal RF front-end case.

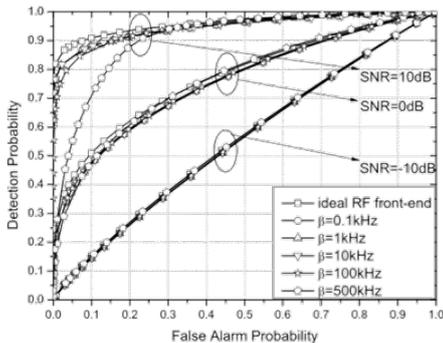

**Fig. 3: Receiver operation curves (ROCs) for different values of β [9].**

*IQI:* It stems from the unavoidable amplitude and phase differences between the physical analog in-phase (I) and quadrature (Q) signal paths at the up- and down-converter of the TX and RX, respectively. In particular, IQI occurs due to the error in the nominally $90^o$ error shifter and the mismatch between the amplitudes of the LO I and Q outputs. This problem arises mainly due to the finite tolerances of the capacitors and the resistors used in the implementation of the analog front-end components. As depicted in Fig. 2.d., in the case of multi-channel spectrum sensing, IQI results in mirror-channel interference, which causes an energy reduction on the occupied channel and a corresponding energy increase in the idle mirror-channel. Fig. 4 numerically quantifies the impact of IQI in the spectrum sensing capabilities of the CR. Again, we assume that the SNR in all the K channels is the same. Fig. 4 indicates that as the level of this imperfection increases, i.e., as the image rejection ratio (IRR) decreases, the interference of the mirror channel increases; hence, the spectrum sensing capability of the low-cost CR decreases.

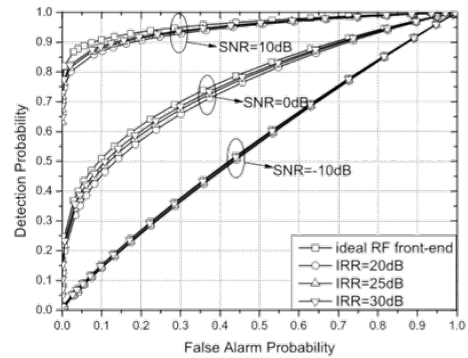

**Fig. 4: ROCs for different values of image rejection ratio (IRR) [9].**

Joint impact of RF impairments in spectrum sensing: The RF imperfections result in not only amplitude/phase distortion, but also neighbor and mirror interference, as demonstrated intuitively in Fig.2.e. amplifier's nonlinearities cause amplitude/phase distortion and an additive nonlinear distortion noise, whereas phase noise causes interference to the received base band signal at the $k-$th channel, due to the received base band signals at the neighbor channels $k-1$ and $k+1$. The joint effects of phase noise and IQI result in interference to the signal at the k-th channel by the signals at the channels -k-1, -k, -k+1, k-1 and k+1. Furthermore, the joint effects of LNA nonlinearties and IQI result in additive distortion noises and mirror channel interference. Fig.2.e. clearly demonstrates that LNA nonlinearities, IQI and phase noise results in an amplitude and phase distortion, as well as interference to channel k

# IEEE COMSOC TCCN Communications

from the channels -k-1, -k, -k+1, k-1 and k+1, plus a distortion noise. If channel k is busy, the received signal's energy at channel k is increased, due to the interference of the neighbor and mirror channels, hence, the decision will be more accurate. However, if channel k is idle, the received signal's energy at channel k, due to the interference and the noise, may be greater than the decision threshold, and the detector will wrongly decide that the channel is busy. Consequently, the interference due to hardware imperfections plays an important role in the spectrum sensing capabilities; therefore, it should be quantified and taken into consideration when selecting the detection threshold.

## 4. Conclusion

In this paper, we presented the impact of hardware impairments in the spectrum sensing performance of low-cost CRs. Both the academia and the industry have several concerns regarding the limitations caused by the hardware imperfections and the self-interference leakage; especially in high-data rate systems. Therefore, several studies were focused on quantifying their impact in spectrum sensing and revealed that RF imperfections can significantly limit the CR performance and capability to identify spectrum holes. On the other hand, spectrum sensing solutions that take into account the impact of hardware imperfections of the CR's RF chain has not yet fully investigated and is a subject for future research. Such solutions are expected to drastically increase the spectrum utilization and deal with the spectrum scarcity issue in a more efficient manner.

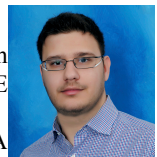

**Dr. Alexandros-Apostolos A. Boulogeorgos:** IEEE Member, Researcher at University of Pire-aus, University of Thessaly and Visiting Lecturer at University of Western Macedonia, Greece.

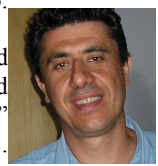

**Prof. George K. Karagiannidis:** IEEE Fellow, Professor at Aristotle University of Thessaloni-ki, Greece. 2015, 2016 and 2017 Highly Cited Researcher.